\documentclass{pasj00}
\begin{document}
\title{Isolated Millimeter Flares of Cyg X-3 }
\SetRunningHead{T\sc{suboi} \rm et al. }{Cyg X-3}

\author{Masato T\sc{suboi} \rm and Kazuhisa K\sc{amegai}}
\affil{Institute of Space and Astronautical Science, Sagamihara, Kanagawa 252-5210}

\author{Atsushi M\sc{iyazaki}}
\affil{Mizusawa VLBI Observatory, National Astronomical Observatory, Mizusawa, Oshu, Iwate 023-0861}
\affil{Korean VLBI Network, Korea Astronomy and Space Science Institute,\\
 Seongsan-ro 262, Seodaemun-gu,
Seoul 120-749, KOREA}

\author{Kouichiro N\sc{akanishi}}
\affil{National Astronomical Observatory, 
Osawa, Mitaka, Tokyo 181-8588 }
\and
\author{Taro K\sc{otani}}
\affil{Waseda University, 17 Kikuicho, Shinjuku, Tokyo 162-0044}

\maketitle

\begin{abstract}
Cygnus X-3 (Cyg X-3) is a well-known microquasar with relativistic jets.  
Cyg X-3 is especially famous for its giant radio outbursts, which have been observed once every few years since their first discovery. 
Each giant outburst presumably consists of a series of short-duration flares. The physical parameters of the flares in the giant outbursts are difficult to derive because the successive flares overlap. 
Here, we report isolated flares in the quiescent phase of Cyg X-3, as observed at 23, 43, and 86 GHz with the 45-m radio telescope at Nobeyama Radio Observatory. The observed flares have small amplitude (0.5--2 Jy) and short duration (1--2 h).  The millimeter fluxes rapidly increase and then exponentially decay. The lifetime of the decay is shorter at higher frequency. The radio spectrum of Cyg X-3 during the flares is flat or inverted around the peak flux density.  After that, the spectrum gradually becomes steeper. The observed characteristics are consistent with those of adiabatic expanding plasma. The brightness temperature of the plasma at the peak is estimated to be $T_B\gtrsim 1 \times 10^{11}$ K. The magnetic field in the plasma is calculated to be $0.2 \lesssim H \lesssim 30$ G.

\end{abstract}
 
\section{Introduction} 
Cygnus X-3 (Cyg X-3) is a well-known microquasar that contains a Wolf--Rayet star and a compact object, the nature and mass of which are still not known (\cite{Lomment2005},\cite{Shrader}). 
Giant radio outbursts from Cyg X-3 have been observed once every few years since their first discovery. 
Very long baseline interferometry observations of these outbursts revealed that Cyg X-3 sometimes has a bipolar relativistic jet (\cite{Marti}, \cite{Miller-Jones2004},  \cite{Tudose}) and sometimes has a one-sided jet (\cite{Mioduszewski}). These relativistic jets are presumably the remnants of ejection from the flares. 
In addition, Cyg X-3 was first detected in the GeV $\gamma$-ray regime in 2009 by the Fermi (\cite{Abdo}) and AGILE (\cite{Tavani}) satellites. 
Williams et al. reported that a radio flare was followed by a $\gamma$-ray flare(\cite{Williams}).
The close relationship between radio and $\gamma$-ray flares is also supported by new observations (\cite{Kotani}, \cite{Bulgarelli}). 
The centimeter wave emission  from Cyg X-3 has been monitored by several groups, even during the quiescent phase (e.g. \cite{Trushkin2008}). The Green Bank interferometer at 5 GHz and the Ryle telescope at 15 GHz have revealed the detailed behavior of radio emission from Cyg X-3 with a high time resolution (e.g. \cite{Ogley2001}). The centimeter radio emission changes gradually over a time scale of 1 h or less. In contrast, the time scale of the change during outbursts is shorter at higher frequency. In particular, the time scale  for millimeter waves is much shorter than that for centimeter waves (e.g. \cite{Tsuboi2008}). 

The giant outbursts of Cyg X-3 are presumed to consist of a series of short-duration flares. In previous research, a millimeter wave monitor with sufficient time coverage and high time resolution found a new transient phenomenon of Cyg X-3: a flux density change with an e-folding rise time of 3.6 min or less in the millimeter light curve (\cite{Tsuboi2010}). However, such flares in the giant outburst are too crowded together to discriminate the physical parameters of a single flare, even in the millimeter wave region. Ascertaining these parameters is vital for exploring the mechanism of the flares. 

Accordingly, to obtain light curves that do not suffer from the interference of successive flares, observations were performed in the millimeter wave region at a high time resolution during  quiescent phase of Cyg X-3 by using the 45-m radio telescope at Nobeyama Radio Observatory\footnote{Nobeyama Radio Observatory is a branch of the National Astronomical Observatory, National Institutes of Natural Sciences, Japan} (NRO45). This telescope has wide frequency coverage from 20 to 100 GHz. Studying the behavior of flares at different frequencies also provided important information for exploring the mechanism of the flares.  

\section{Observations}
We monitored the flux density of Cyg X-3 in the quiescent phase at a high time resolution (4--10 min) by using NRO45 at 23, 43, and 86 GHz for 20 days during the period from 2006 to 2010. 
 
 To conduct the observations, a cooled high electron mobility transistor receiver with a dual circular polarization feed was used at 23.0 GHz. However, observations at 23 GHz were performed only in 2006 because changing the frequency between 23 GHz and 43 and 86 GHz required a dead time of 5 min, which we judged to be a long time when compared with the time scale of Cyg X-3 phenomena. Superconductor-insulator-superconductor receivers with orthogonal linear polarization feeds were used to make observations at 43.0 and 86.0 GHz. The frequency bandwidth of these receivers is 500 MHz. The system noise temperatures during the observations, including atmospheric effects and antenna ohmic loss, were 80--120 K at 23 GHz, 120--200 K at 43 GHz, and 250--350 K at 86 GHz. The full width at half-maximum values of the telescope beams were $77"$ at 23 GHz, $39"$ at 43 GHz, and $19"$ at 86 GHz. 
 To remove any atmospheric effects, the telescope beam was alternated at 15 Hz between the position of the source and a region of blank sky 6'  from the source in the azimuth direction by a beam switch.
The telescope was position-switched between an on-source position at $\Delta Az/\cos El=0'$ and a region of blank sky at $\Delta Az/\cos El=3'$, where the duration at each position was 8 s/cycle. 
The antenna temperatures were calibrated by a chopper wheel method. The primary flux calibrator for the conversion from antenna temperature to flux density was a proto-planetary nebula, NGC 7027, whose flux density values are given as 5.5 Jy at 22 GHz, 5.0 Jy at 43 GHz, and 4.6 Jy at 86 GHz (\cite{Ott}). Telescope pointing was checked and corrected in every observation procedure by viewing NGC 7027 in cross-scan mode. The pointing accuracy
was better than $3"$ root mean square  during the observations. 

The uncertainty in the flux density of Cyg X-3 depends on the weather conditions. However, the sensitivity of the telescopes is not a principal factor in the uncertainty. Because the primary flux-scale calibrator, NGC 7027, is close to Cyg X-3 in the celestial sphere and the difference in atmospheric attenuation between these sources is calibrated by the chopper wheel method, there is no significant effect on the flux density data. The typical systematic uncertainty for NRO45 is 10\%; however, the relative uncertainties in the flux density on any particular day should be much less than this value because the observations were performed under fine weather conditions.

 \section{Results}
 From the 20 days of monitoring during the quiescent phase of Cyg X-3, we found two isolated flares, at MJD $=53809$ (15 March 2006 UT) and at MJD $=54972$ (May 20 2009 UT). We observed the whole episode of these flares without contamination from successive flares. There was no isolated flare with statistical significance on other days. The complete monitoring data will be published in a future paper.

 \subsection{Flare at MJD $=53809$}
Figure 1a shows the light curves of  the flare at MJD$53809$ at 23, 43, and 86 GHz. 
The MJD $=53809$ flare was found in the post-bursting phase of the giant outburst in spring 2006. The flux densities rapidly increased to a peak around MJD $=53809.78$. Unfortunately, because calibration was performed near the peak, there is a break in the data. The maximum flux densities were at least $S_\nu=1.17\pm0.01$ Jy at 43 GHz and $S_\nu=1.71\pm0.05$ Jy at 86 GHz, where the error includes only statistical error. 
The time scales of the e-folding rise, $t_e$, were below $t_e\lesssim7$ min at 43 GHz and $t_e\lesssim6$ min at 86 GHz. In an earlier paper (\cite{Tsuboi2010}), we reported an abrupt large-amplitude change in flux density with an e-folding rise time of 4 min or less in light curves at 100 GHz (\cite{Tsuboi2010}). The time scales shown in Figure 1a are similar to the previous value. In contrast, the flux density  appears to change more gradually at 23 GHz than at 43 GHz and 86 GHz, suggesting that the time scale becomes longer at lower frequency. However, the time lag between 43 and 86 GHz is not clear. The absence of time lag is similar to that seen for light curves at 100 GHz (\cite{Tsuboi2010}). 
After the intensity peak, the flux densities then exponentially decayed. The lifetimes ($S_\nu\propto\exp{(-(t-t_0)/\tau)}$) of the fluxes were $\tau=0.167_{-0.021}^{+0.027}$ days at 23 GHz (dotted line in Figure 1a), $\tau=0.098_{-0.015}^{+0.021}$ days at 43 GHz (broken line in Figure 1a), and $\tau=0.033_{-0.005}^{+0.007}$ days at 86 GHz (solid line in Figure 1a).  
Here, the initial time of the flare, $t_0$, is assumed to be the same as the time at the peak flux density, $t_0=53809.78$ in MJD. The lifetime of the flux was much shorter at higher frequency. In addition, there might be a plateau around MJD $=53809.83$--$53809.86$ in the flux density at 86 GHz. A plateau was not clearly seen in the flux densities at 23 and 43 GHz. 

Figure 2a shows the radio spectral index of Cyg X-3 at MJD $=53809$. The values are derived from the inclination between the flux densities at 43 and 86 GHz. The radio spectrum of Cyg X-3 is inverted, $\alpha\sim-0.5$($S_\nu\propto\nu^{-\alpha}$), around the peak flux density. 
After the peak flux density, the spectral index rapidly becomes steeper; however, during the exponential decay of the flux density, the spectral index is around $\alpha\sim0.5$.

 \subsection{Flare at MJD $=54972$}
 Figure 1b shows the light curves of the flare at MJD $=54972$ at 43 and 86 GHz. MJD $=54972$ corresponded to the quiescent phase of Cyg X-3 in spring 2009. This flare has been briefly reported (\cite{Tsuboi2009}). The flux densities rapidly increased to a peak around MJD $=54972.88$. The maximum flux densities are $S_\nu=0.43\pm0.02$ Jy at 43 GHz and $S_\nu=0.39\pm0.02$ Jy at 86 GHz. The time scales of the e-folding rise are $t_e\lesssim11$ min at 43 GHz and $t_e\lesssim7$ min at 86 GHz, slightly longer than those in Figure 1a.
The flux densities exponentially decayed after the intensity peaked.  The lifetimes of the fluxes, assuming $t_0=54972.88$, are $\tau=0.077_{-0.003}^{+0.004}$ days at 43 GHz (broken line in Figure 1b), and $\tau=0.033_{-0.007}^{+0.012}$ days at 86 GHz (solid line in Figure 1b). These lifetimes are consistent with those in Figure 1a. There might also be a plateau in the flux density at 86 GHz around MJD $=54972.91$--$54972.95$. In contrast to the first flare, a similar feature is also observed at 43 GHz. 

Figure 2b shows the radio spectral index of Cyg X-3 at MJD $=54972$. The radio spectrum of Cyg X-3 is flat around the peak flux density, but rapidly become steeper to $\alpha\sim0.5$ during the exponential decay of the flux density. This is consistent with the spectral index in Figure 2a.

\clearpage
\begin{figure}
\begin{center}
\includegraphics[width=12cm]{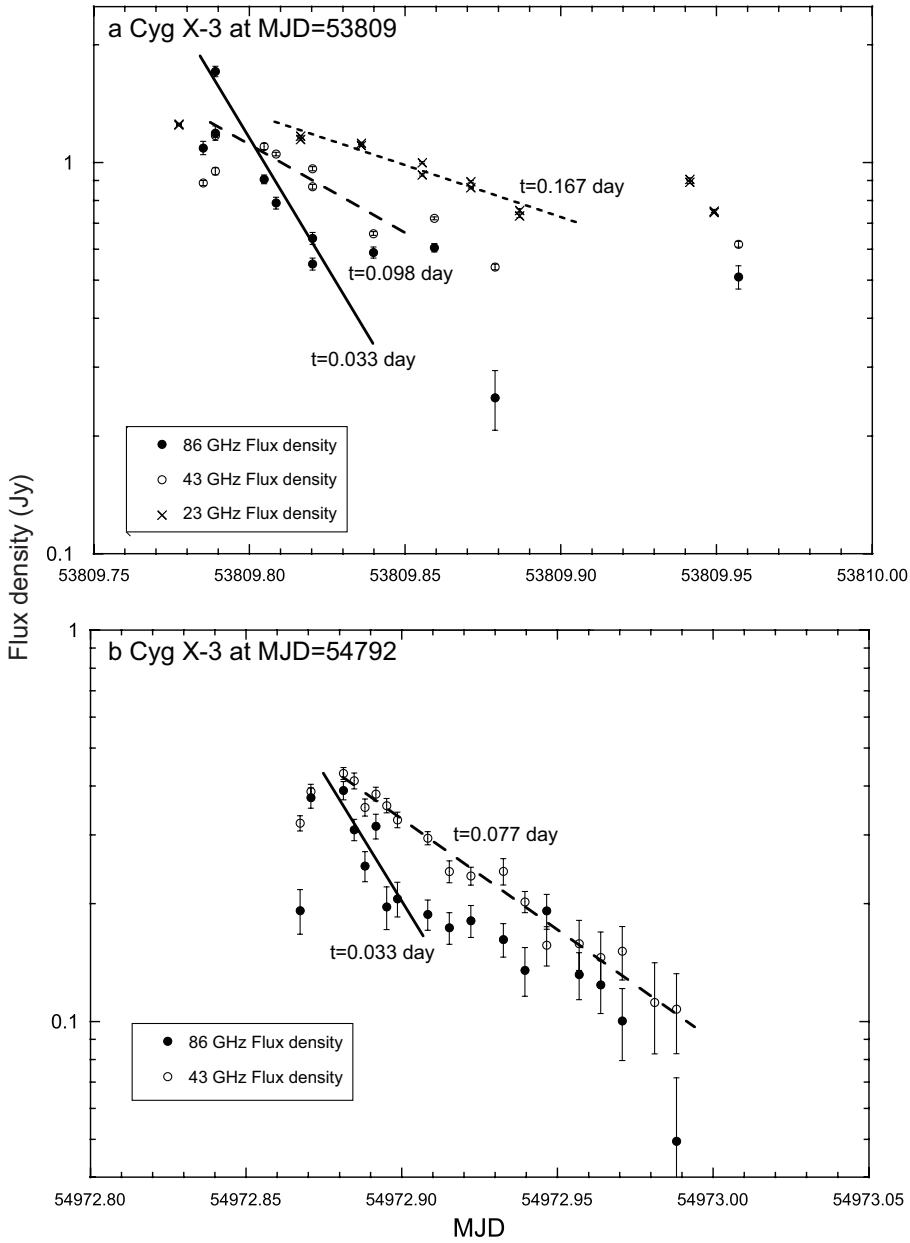}
\caption{\textbf{a} Light curves of Cyg X-3 at MJD $=53809$ observed with the NRO45 telescope. The frequency bands are 23 (cross), 43 (open circle), and 86 (filled circle) GHz. The maximum flux densities occur around MJD $=53809.78$. They are $S_\nu=1.17\pm0.01$ Jy at 43 GHz and $S_\nu=1.71\pm0.05$ Jy at 86 GHz. After the intensity peak, the flux densities exponentially decayed. The lifetimes ($S\nu\propto\exp{(-(t-t_0)/\tau)}$) are $\tau=0.167_{-0.021}^{+0.027}$ days at 23 GHz (dotted line), $\tau=0.098_{-0.015}^{+0.021}$ days at 43 GHz (broken line), and $\tau=0.033_{-0.005}^{+0.007}$ days at 86 GHz (solid line). Here, the initial time of the flare, $t_0$, is assumed to be the time at the peak, $t_0= 53809.78$.
\textbf{b} Light curves of Cyg X-3  at MJD $=54972$ observed with the NRO45 telescope. The frequency bands are 43 (open circle) and 86 (filled circle) GHz. The maximum flux densities occur around MJD $=54972.88$. They are $S_\nu=0.43\pm0.02$ Jy at 43 GHz and $S_\nu=0.39\pm0.02$ Jy at 86 GHz. The lifetimes, assuming $t_0=54972.88$, are $\tau=0.077_{-0.003}^{+0.004}$ days at 43 GHz (broken line), and $\tau=0.033_{-0.007}^{+0.012}$ days at 86 GHz (solid line).
}
\label{Fig1}
\end{center}
\end{figure}
\clearpage

\clearpage
\begin{figure}
\begin{center}
\includegraphics[width=16cm]{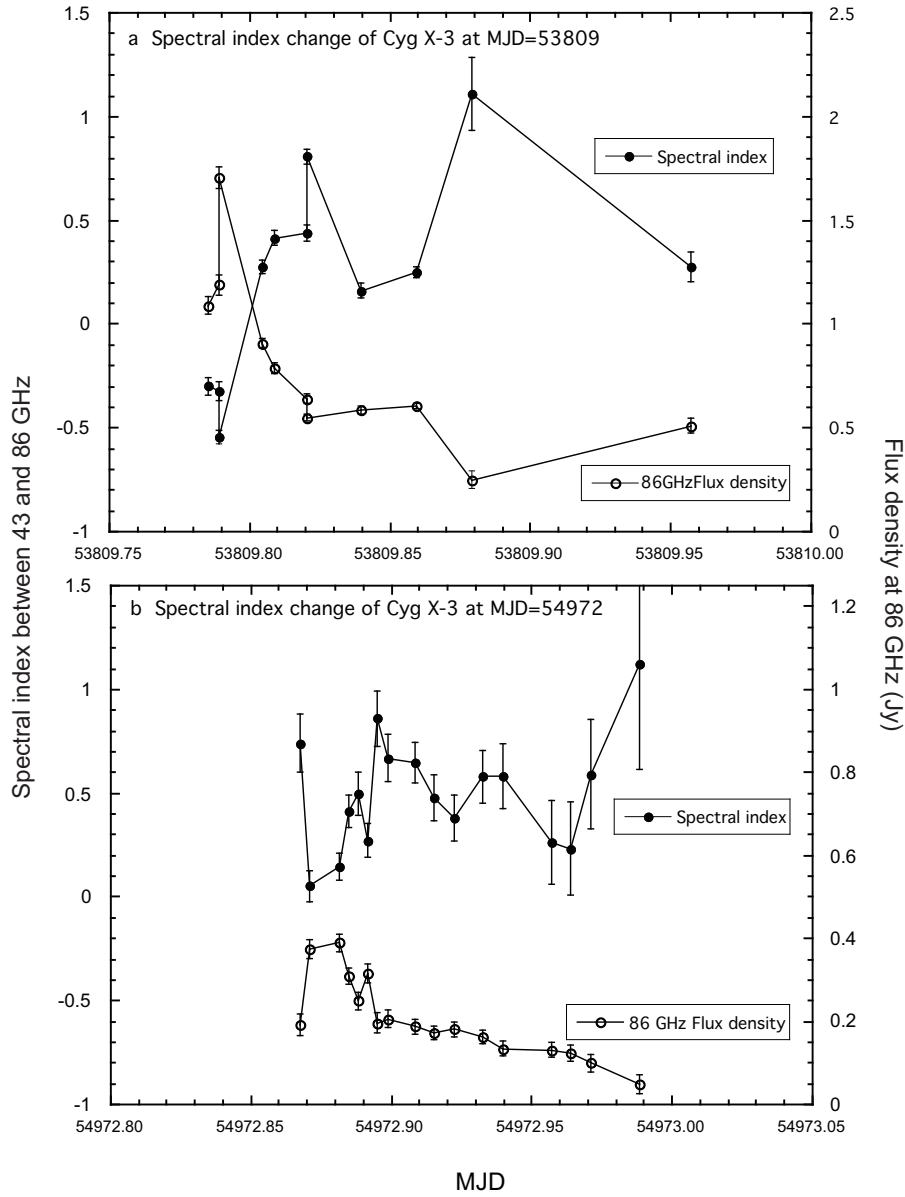}
\caption{ \textbf{a} Spectral index between 43 and 86 GHz (filled circle) of Cyg X-3  at MJD $=53809$ observed with the NRO45 telescope. The flux density at 86 GHz (open circle) is also shown for comparison. The radio spectrum of Cyg X-3  is inverted, $\alpha\sim-0.5$($S_\nu\propto\nu^{-\alpha}$), around the peak flux density. The spectral index was around $\alpha\sim0.5$ during the exponential decay of the flux density.
\textbf{b} Spectral index between 43 and 86 GHz (filled circle) of Cyg X-3 at MJD $=54972$ observed with the NRO45 telescope. The flux density at 86 GHz (open circle) is also shown for comparison. The radio spectrum of Cyg X-3 has a flat spectrum around the peak flux density. The spectral index was at around $\alpha\sim0.5$ during the exponential decay of the flux.
}
\label{Fig1}
\end{center}
\end{figure}
\clearpage

\section{Discussion} 
The time lag between 43 and 86 GHz was not observed in the rising phase of the flares; the two flares had flat or inverted spectra around the maximum intensity. The flux densities of both flares exponentially decayed over a similar lifetime and with a similar spectral index after the peak. These observed characteristics are consistent with those of adiabatic expanding plasma; that is, an optically thick plasma blob is ejected. This blob rapidly expands and its flux density increases rapidly, independent of frequency. The optical thickness decreases with the expansion and the plasma blob becomes optically thin at a higher frequency.  The magnetic field producing the synchrotron emission decays. The emission at higher frequency decays more quickly. 

The radius of the emitting region at the maximum intensity, $r$, can be estimated from $r< ct_e/2$, and the upper limit of the source size is $r< 0.4$ AU. The observation frequency is $\nu = 86$ GHz and the flux density is $S_{\nu}= 1.7$ Jy for the flare at MJD $=53809$. 
The brightness temperature of the emitting region, $T_\mathrm{B}$, is given by
\begin{equation}
\label{ }
T_\mathrm{B}=\frac{c^2}{2k_B\nu^2}S_{\nu}\Bigl(\frac{d}{r}\Bigr)^2,
\end{equation}
where $d$ is the distance of Cyg X-3 and $k_B$ is Boltzmann conatatnt (\cite{Ogley2001}). This brightness temperature is larger than
$T_\mathrm{B}>1.5\times10^{11}$ K, assuming that Cyg X-3 is located at a distance of $d=9$ kpc. 

The observed spectrum of the flares around the peak is flat or inverted between $\nu=43$ and $86$ GHz. If the emission is completely optically thick in the frequency range, the spectral index should be $\alpha\sim-2.5$ for relativistic electrons with a power energy spectrum, $N(E)\propto E^{-\gamma}$. The observed spectral indexes indicate that emission in this frequency range is close to the turnover frequency, at which the optical thickness becomes $\tau\sim 1$ and is given by
\begin{equation}
\label{ }
\nu_s^{\frac{\gamma+4}{2}} \sim c_{14}(\gamma)\frac{S_{\nu}\nu^{\frac{\gamma-1}{2}}}{\pi r^2/d^2}H_{\bot}^{\frac{1}{2}},
\end{equation}
(see Eq. 6.38 in \cite{Pacholczyk}).
 
We assume that the electron energy index is $\gamma= 2$, which is derived from the spectral index of $\alpha\sim0.5$ during the exponential decay. Hence, with $\gamma= 2$ the turnover frequency becomes 
\begin{equation}
\label{ }
\nu_s[\mathrm{GHz}]\sim 42\times H_{\bot}[\mathrm{G}]^{\frac{1}{5}}.
\end{equation} 
Here, we use the coefficient $c_{14}(2) \sim 3.42\times 10^{30}$. 
On the other hand, the turnover frequency in the flares is estimated to be $\nu_s\sim 80$ GHz, owing to the observed flares showing a flat or inverted spectrum between $\nu=43$ and $86$ GHz. The magnetic field strength at the peak is calculated to be $H\sim 30$ G and the cooling time of the electrons, $t_c$, is estimated to be
\begin{equation}
\label{ }
t_c[s]\sim \frac{1}{2}c_{12}\times H_{\bot}[\mathrm{G}]^{-\frac{3}{2}},
\end{equation} 
where $c_{12}=1.6\times 10^{7}$ (see Eq. 7.7 in \cite{Pacholczyk}). If the magnetic field is $H_{\bot}\lesssim 30$ G, the cooling time is $t_c\gtrsim14$ h, which is consistent with no change occurring in the spectral index during the observed exponential decay. The exponential decay is assumed to be caused by the decrease in the strength of the magnetic field. 

Assuming equipartition between the magnetic field and the electrons, the minimum energy of the source to account for the observed synchrotron
emission is approximated by
\begin{eqnarray}
 E_{\rm min} &\sim &3\times 10^{33}\left(1+\frac {\epsilon_{\rm p}}{\epsilon_{\rm e}}\right)^{4/7} 
  \left ( \frac{t_{\rm e}}{\rm s} \right )^{9/7}
  \left ( \frac{\nu}{\rm GHz} \right )^{2/7}
  \left ( \frac{S_\nu}{\rm mJy} \right )^{4/7}
  \left ( \frac{d}{\rm kpc} \right )^{8/7}\;\;{\rm erg}\\
&= &2.1 \times 10^{40} \;\;{\rm erg},
\end{eqnarray}
where $\epsilon_{\rm p}/\epsilon_{\rm e}$ is the ratio between the energy in protons and that in electrons; this ratio is assumed to be zero hereafter \citep{Fender2003}.  The minimum power given to the electrons is found by dividing the minimum energy by the time scale of the e-folding rise $t_{\rm e}$:
\begin{equation}
 P_{\rm min} \sim \frac{ E_{\rm min}}{t_{\rm e}} = 5.1\times10^{37}\;\; \rm erg\; s^{-1}.
\end{equation}
The power is comparable with those estimated from flux densities at low frequencies, whose rise time scales are longer.

From the assumption of equipartition, the magnetic field strength, $H_{\rm eq}$, is
\begin{equation}
H_{\rm eq} \sim 30 \times   
\left ( \frac{S_\nu}{\rm mJy} \right )^{2/7}
  \left ( \frac{d}{\rm kpc} \right )^{4/7}
  \left ( \frac{t_{\rm e}}{\rm s} \right )^{-6/7}
  \left ( \frac{\nu}{\rm GHz} \right )^{1/7}\;\;{\rm G} = 10\;\;{\rm G}. 
\end{equation}
Note that the equipartition magnetic field strength, $H_{\rm eq} \sim 10$ G, is of the same order as the strength of $H \sim 30 $ G, obtained from the turnover frequency.

The ratio of the inverse-Compton scattering loss, $\Bigl(\frac{dE}{dt}\Bigr)_C$, and the synchrotron radiation loss, $\Bigl(\frac{dE}{dt}\Bigr)_S$, for the relativistic electrons is given by
\begin{equation}
\label{ }
R=\frac{\Bigl(\frac{dE}{dt}\Bigr)_C}{\Bigl(\frac{dE}{dt}\Bigr)_S}
=\frac{u_{rad}}{u_{mag}} = \frac{\frac{\nu S_{\nu}}{c}\Bigl(\frac{d}{r}\Bigr)^2}{\frac{H^2}{8\pi}},
\end{equation}
where $u_{rad}$ and $u_{mag}$ are the radiation energy density and the magnetic energy density, respectively (see Eq. 5.60 in \cite{Pacholczyk}, \cite{Tsuboi2010}). 
If the inverse-Compton scattering loss dominates the synchrotron radiation loss, the lifetime of the relativistic electron would be so short that the emission would be invisible.
Therefore, we set the ratio $R\le1$ by assuming that the spectral index is flat until $\nu\sim100$ GHz. In this case, the magnetic field is greater than $H\ge 0.2$ G and thus, the magnetic field strength of Cyg X-3 is estimated to be in the range $0.2 \lesssim H \lesssim 30$ G.

\section{Conclusions} 
We performed millimeter wave monitor observations of Cyg X-3 in quiescent phases between giant outbursts at 23, 43, and 86 GHz by using the NRO45 radio telescope. We found two isolated flares with amplitude 0.5--2 Jy, for which the light curves did not suffer from successive flares. The e-folding rise times were as short as 6--7 min at 86 GHz. 
The spectra were flat or slightly inverted around the maximum intensity. The flux densities of both flares exponentially decayed over a similar lifetime, $\tau=0.03$ days at 86 GHz, and with a similar spectral index, $\alpha\sim0.5$. The observed characteristics are consistent with those of adiabatic expanding plasma. 
The observed e-folding rise times showed that the upper limit radius of the ejecting plasma was smaller than $r<0.4$ AU. The brightness temperature of the ejecting plasma was estimated to be $T_B\gtrsim 1 \times 10^{11}$ K. The magnetic field in the plasma was derived to be $0.2 \lesssim H \lesssim 30$ G.

\bigskip 
The authors thank the members of the 45-m radio telescope group at Nobeyama Radio Observatory for support in the observations. This research was supported by KAKENHI 19047002 (TK) and by KAKENHI 23340075 (TK).

\end{document}